\documentclass[11pt]{article}
\usepackage[dvips]{graphicx}
\begin{document}
\begin{center}
{\large\bf Quantum Cosmology and Dark Energy Model of Born-Infeld Type Scalar Field}\\
 \vskip 0.15 in
H.Q.Lu$^1$  Z.G.Huang$^2$  W.Fang$^1$ P.Y. Ji$^1$\\
$^1Department~of~Physics,~Shanghai~University,$\\
$~Shanghai,~200444,~P.R.China$\\
$Email:alberthq_-lu@staff.shu.edu.cn$\\
$^2Department~of Mathematics~and~Science,~Huaihai~Institute~of~Technology,$\\
$~Lianyungang,~222005,~P.R.China$\\
 \vskip 0.7 in
\centerline{\textbf{Abstract}}
\end{center}
{\small ~~In this paper, we consider a quantum model of gravitation
interacting with a Born-Infeld(B-I) type scalar field $\varphi$. The
corresponding Wheeler-Dewitt equation can be solved analytically for
both very large and small cosmological scale factor. In the
condition that small cosmological scale factor tend to limit, the
wave function of the universe can be obtained by applying the
methods developed by Vilenkin, Hartle and Hawking. Both Vilenkin's
and Hartle-Hawking's wave function predicts nonzero cosmological
constant. The Vilenkin's wave function predicts a universe with a
 cosmological constant as large as possible, while the
Hartle-Hawking's wave function predicts a universe with positive
cosmological constant, which equals to $\frac{1}{\lambda}$. It is
different from Coleman's result that cosmological constant is zero,
and also different from Hawking's prediction of zero cosmological
constant in quantum cosmology with linear scalar field. We suggest
that dark energy in the universe might result from the B-I type
scalar field with potential and the universe can undergo a phase of
accelerating expansion. The equation of state parameter lies in the
range of $-1<$w$<-\frac{1}{3}$. When the potential
$V(\varphi)=\frac{1}{\lambda}$, our Lagrangian describes the
Chaplygin gas. In order to give a explanation to the observational
results of state parameter w$<-1$, we also investigate the phantom
model that posses negative kinetic energy. We find that weak and
strong energy conditions are violated for phantom B-I type scalar
field. At last, we study a specific potential with the form
$V_0(1+\frac{\varphi}{\varphi_0})e^{-(\frac{\varphi}{\varphi_0})}$
in phantom B-I scalar field in detail. The
attractor property of the system is shown by numerical analysis.
}

\vskip 0.3 in \begin{center}\textbf{I.INTRODUCTION}\end{center}
~~~~Today, one of the central questions in cosmology is the origin
of the dark energy. Many candidates for dark energy have been
proposed. Among these models, the important ones are cosmological
constant, "quintessence"[7], "K-essence"[8] and "tachyon field"[9].
On the other hand, nonlinear Born-Infeld theory has been considered
in string theory and cosmology.
\par In this paper, we consider the quantum cosmology  and
dark energy model of nonlinear Born-Infeld type scalar field. The
corresponding Lagrangian has been first proposed by Heisenberg[1]
to describe the process of meson multiple production connected
with strong field regime, as a generalization of the B-I
electromagnetic field Lagrangian
$L^{BI}=b^2[1-\sqrt{1+(\frac{1}{2b^2})F_{\mu\nu}F^{\mu\nu}}]$[2].
The Lagrangian of B-I type scalar field is
\begin{equation}L_s=\frac{1}{\lambda}(1-\sqrt{1-\lambda\varphi_{,\mu}\varphi_{,\nu}g_{\mu\nu}})-V(\varphi)\end{equation}
Eq.(1) posses some interesting characteristics[3] that nonsingular
scalar field solution can be generated, and shock waves don't
develop under smooth
 and continuous initial conditions. Second when
$g^{\mu\nu}\varphi_{,\mu}\phi_{,\nu}\ll\frac{1}{\lambda}$, by
Taylor expansion, Eq.(1) approximates to
\begin{equation}L_s=\frac{1}{2}g^{\mu\nu}\varphi_{,\mu}\phi_{,\nu}-V(\varphi)\end{equation}
the linear theory is recovered. Third, when potential
$V=\frac{1}{\lambda}$, our Lagrangian represents a universe filled
with chaplygin gas which maybe unify dark energy and dark matter[5].
\par We consider quantum creation of universe based on the
Wheeler-Dewitt(WD) equation $\hat{H}\dot{\psi}=0$ in the superspace.
This quantum approach applying to cosmology may help us avoid the
cosmology singularity problem and understand what determined the
initial state of the universe. In section II, we consider quantum
cosmology with B-I type scaler field and constant potential
$V(\varphi)$ which describes cosmological constant effectively. We
also find the wave function of universe by applying the methods
developed by Vilenkin and Hartle-Hawking[4]. Both Vilenkin's and
Hartle-Hawking's wave function predicts nonzero positive
cosmological constant. It is different from zero cosmological
constant that has been predicted by Coleman[10] and Hawking[4]. In
fact the H-H's wave function with B-I type scalar field predicts a
universe filled with Chaplygin gas. In section III, we investigate
the cosmology with B-I type scalar field as dark energy.  Section IV
is conclusion. \vskip 0.3 in
\begin{center}\textbf{II.QUANTUM COSMOLOGY WITH B-I TYPE SCALAR FIELD}\end{center}
\par In order to find the solution of the WD equation, we shall
apply the minisuperspace model-- a closed Robertson-Walker(R-W)
metric. In the minisuperspace there are only two degrees of
freedom: the scale factor $a(t)$ and scalar field $\varphi(t)$.
Using Eq.(1) and by integrating with respect to space-components,
the action $S=\int\frac{R}{16\pi G}\sqrt{-g}d^4x+\int
L_s\sqrt{-g}d^4x$ becomes
\begin{equation}S=\int\frac{3\pi}{4G}(1-\dot{a}^2)adt+\int2\pi^2a^3[\frac{1}{\lambda}(1-\sqrt{1-\lambda\dot{\varphi}^2})-V(\varphi)]dt=\int {\mathcal{L}}_gdt+\int {\mathcal{L}}_sdt\end{equation}
where the upper-dot means the derivative with respect the time
$t$. From the Euler-Lagrange equation
\begin{equation}\frac{d}{dt}(\frac{\partial {\mathcal{L}}_s}{\partial\dot{\varphi}})-\frac{\partial {\mathcal{L}}_s}{\partial\varphi}=0\end{equation}
we can obtain
\begin{equation}\dot{\varphi}=\frac{c}{\sqrt{a^6+\lambda c^2}}\end{equation}
where $c$ is integral constant. From the above equation we know
that cosmological scale factor $a$ is very large or small when
$\dot{\varphi}$ is very small or large respectively. The critical
kinetic energy $\frac{1}{2}(\dot{\varphi})_{max}^2$ is
$\frac{1}{2\lambda}$.
\par To quantize the model, we first find
out the canonical momenta $P_a=\partial
{\mathcal{L}}_g/\partial\dot{a}=-(3/2G)a\dot{a}$,
$P_\varphi=\partial
{\mathcal{L}}_\varphi/\partial\dot{\varphi}=2\pi^2a^3\dot{\varphi}/\sqrt{1-\lambda\dot{\varphi}^2}$
and the Hamiltonian
$H=P_a\dot{a}+P_\varphi\dot{\varphi}-{\mathcal{L}}_g-{\mathcal{L}}_s$.
$H$ can be written as the follows
\begin{equation}H=-\frac{G}{3\pi a}P_a^2-\frac{3\pi}{4G}a[1-\frac{8\pi G}{3}a^2 V(\varphi)]-\frac{2\pi^2a^3}{\lambda}[1-\sqrt{1+\frac{\lambda P^2_\varphi}{4\pi^4a^6}}]\end{equation}
For $\dot{\varphi}^2\ll\frac{1}{\lambda}$, the Hamiltonian Eq.(6)
can be simplified by using the Taylor expansion, and the terms
smaller than $\dot{\varphi}^6$ can be ignored, so the Hamiltonian
becomes
\begin{equation}H=-\frac{G}{3\pi a}P_a^2-\frac{3\pi}{4G}a[1-\frac{8\pi G}{3}a^2 V(\varphi)]+\frac{P_\varphi^2}{4\pi^2a^3}-\frac{\lambda P_\varphi^4}{64\pi^6a^9}\end{equation}
If  $\dot{\varphi}(\lambda\dot{\varphi}^2\sim1)$ is very large,
Eq.(6) becomes
\begin{equation}H=-\frac{G}{3\pi a}P_a^2-\frac{3\pi}{4G}a\{1-\frac{8\pi G}{3}a^2[ V(\varphi)-\frac{1}{\lambda}]\}\end{equation}
The WD equation is obtained from $\hat{H}\psi=0$, Eqs.(7) and (8)
by replacing $P_a\rightarrow-i(\partial/\partial a)$ and
$P_\varphi\rightarrow -i(\partial/\partial\varphi)$. Then we
obtain
\begin{equation}[\frac{\partial^2}{\partial a^2}+\frac{p}{a}\frac{\partial}{\partial a}-\frac{1}{a^2}\frac{\partial^2}{\partial\tilde{\Phi}^2}-\frac{\lambda}{16\pi^4a^8}\frac{\partial^4}{\partial\tilde{\Phi}^4}-U(a,\tilde{\Phi})]\psi=0\end{equation}
and
\begin{equation}[\frac{\partial^2}{\partial a^2}+\frac{p}{a}\frac{\partial}{\partial a}-u(a,\tilde{\Phi})]\psi=0\end{equation}
where $\tilde{\Phi}^2=4\pi G\varphi^2/3$ and the parameter $p$
represent the ambiguity in the ordering of factor $a$ and
$\partial/\partial a$ in the first term of Eqs.(7) and (8). We
have also denoted
\begin{equation}U(a,\tilde{\Phi})=(\frac{3\pi}{2G})^2a^2[1-\frac{8\pi G}{3}a^2V(\tilde{\Phi})]\end{equation}
\begin{equation}u(a,\tilde{\Phi})=(\frac{3\pi}{2G})^2a^2\{1-\frac{8\pi G}{3}a^2[V(\tilde{\Phi})-\frac{1}{\lambda}]\}\end{equation}
Eqs.(9) and (10) are the WD equations corresponding to the action
(3) in the case of small and large $\dot{\varphi}$ respectively.
\par
Now we take the ambiguity of the ordering factor $p=-1$ and set
the transformation $(a/a_0)^2=\sigma$, with $a_0$ being the
Plank's length. Taking the Plank constant $h=1$ and the speed of
light $c=1$, $a_0\sim\sqrt{4G/3\pi}$, we obtain from Eq.(9)
\begin{equation}\frac{\partial^2\psi}{\partial\sigma^2}-\frac{1}{\sigma^2}\frac{\partial^2\psi}{\partial\tilde{\Phi}^2}-\frac{\lambda}{16\pi^4a^6_0\sigma^5}\frac{\partial^4\psi}{\partial\tilde{\Phi}^4}-\tilde{U}\psi=0\end{equation}
where $\tilde{U}=(3\pi/4G)^2a_0^4(1-\alpha a^2_0\sigma)$ and
$\alpha=(8\pi G/3)$V. Denoting $m=(\frac{\lambda}{16\pi^4a^6_0})$
and assuming $\psi(\sigma,\tilde{\Phi})\sim
Q(\sigma)e^{-K\tilde{\Phi}}$ with $K$ being an arbitrary constant,
we can take the Eq.(13) as follows
\begin{equation}\frac{d^2Q}{d\sigma^2}-(\frac{K^2}{\sigma^2}+\frac{mK^4}{\sigma^5}+\tilde{U})Q=0\end{equation}
If $a$(or $\sigma$) is large, $\alpha a^2=(8\pi
GV(\varphi)/3)a^2_0\sigma\gg1$ and Eq.(14) approximates to
\begin{equation}\frac{d^2Q}{d\sigma^2}+\hat{\mu}a^6_0\sigma Q=0\end{equation}
where $\hat{\mu}=(3\pi/4G)\alpha$. The general solution of Eq.(15)
can be expressed in terms of Bessel function and is given by
\begin{equation}Q(\sigma)=\sqrt{\sigma}Z_{\frac{1}{3}}(\frac{2\hat{\mu}a^3_0}{3}\sigma^{3/2})\end{equation}
$Q(\sigma)$ is an oscillatory function.
\par In the next step, we
consider the solution of the WD Eq.(10) with $\dot{\varphi}$ being
large. We still keep the ambiguity of the ordering of factor
$p=-1$. By the same transformation $(a/a_0)^2=\sigma$, Eq.10)
becomes
\begin{equation}\frac{d^2\psi}{d\sigma^2}-(\frac{3\pi}{4G})a_0^4(1-H^2a^2_0\sigma)\psi=0\end{equation}
where $H^2=(8\pi G/3)[V(\varphi)-\frac{1}{\lambda}]$. When
$H^2a^2\ll1$, Eq.(17) approximates to
\begin{equation}\frac{d^2\psi}{d\sigma^2}-(\frac{3\pi}{4G})^2a_0^4\psi=0\end{equation}
Solving Eq.(18), we get
\begin{equation}\psi=Ne^{-\frac{3\pi}{4G}a_0^2\sigma}=Ne^{-\frac{3\pi}{4G}a^2}\end{equation}
where $N$ is a constant.
\par We can see that solution(19) is
consist with Vilenkin's tunnelling wave function Eq.(29). When $a$
is larger, $H^2a^2=H^2a_0^2\sigma\gg1$, and Eq.(17) approximates
to
\begin{equation}\frac{d^2\psi}{d\sigma^2}+(\frac{3\pi}{4G})^2a_0^6H^2\sigma\psi=0\end{equation}
Its solution is
\begin{equation}\psi=\sqrt{\sigma}Z_{\frac{1}{3}}(\frac{2\tilde{H}^2a_0^3}{3}\sigma^{\frac{3}{2}})\end{equation}
where $\tilde{H}=(\frac{3\pi}{4G})H$. The wave function $\psi$
given by Eq.(21) is an oscillatory function.
\par Next we will consider the cosmology in case of very large
$\dot{\varphi}$(correspondingly very small $a(t)$) by Vilenkin's
quantum tunnelling approach. Eq.(10) has the form of a
one-dimensional Schr\"{o}dinger equation for a "particle" described
by a coordinate $a(t)$, which is zero energy and moves in a
potential $u$. The classically allowed region is $u\leq0$ or $a\geq
H^{-1}$, with $H$ being defined in Eq.(17). In this region,
disregarding the pre-exponential factor, the WKB solutions of
Eq.(10) are
\begin{equation}\psi_{\pm}^{(1)}(a)=exp\{[\pm i\int_{H^{-1}}^aP(a')da']\mp\frac{i\pi}{4}\}\end{equation}
The under-barrier($a<H^{-1}$, classically forbidden or Euclidean
region) solutions are
\begin{equation}\psi_{\pm}^{(2)}(a)=exp\{[\pm\int^{H^{-1}}_a|P(a')|da']\}\end{equation}
where $P(a)\equiv\sqrt{|-u(a)|}$.
\\The classical momentum conjugate to $a$ is $P_a=-a\dot{a}$. For
$a>H^{-1}$, we have
\begin{equation}(-i\frac{d}{da})\psi^{(1)}_{\pm}(a)=\pm P(a)\psi^{(1)}_{\pm}(a)\end{equation}
and thus $\psi^{(1)}_-(a)$ and $\psi^{(1)}_+(a)$ describe the
expanding and contracting universe respectively. The tunneling
boundary condition requires that only the expanding component should
be present at large $a$,
\begin{equation}\psi_T(a>H^{-1})=\psi_-^{(1)}(a)\end{equation}
The under-barrier wave function is found from WKB connection
formula
\begin{equation}\psi_T(a<H^{-1})=\psi_+^{(2)}(a)-\frac{i}{2}\psi^{(2)}_-(a)\end{equation}
The increasing exponential $\psi_-^{(2)}(a)$ and the decreasing
exponential $\psi_+^{(2)}(a)$ have comparable amplitudes at the
nucleation point $a=H^{-1}$, but away from that point the decreasing
exponential dominates
\begin{equation}\psi_T(a<H^{-1})\approx \psi_+^{(2)}(a)=exp[\frac{\pi}{2GH^2}(1-H^2a^2)^{\frac{3}{2}}]\end{equation}
The "tunneling amplitude"(probability distribution for the initial
values of $V$ in nucleating universe) is
\begin{equation}\frac{\psi_T{(H^{-1})}}{\psi_T(0)}=e^{-\frac{\pi}{2GH^2}}\end{equation}
From Eq.(28) we obtain the result that the tunneling wave function
predicts a nucleating universe with the vacuum energy(i.e
cosmological constant) as large as possible and critical kinetic
energy as small as possible (defining the kinetic energy when $a=0$
as critical kinetic energy, which equals $\frac{1}{2\lambda}$ from
Eq.(5)). If $H^2a^2\ll1$, by Taylor expasion Eq.(27) becomes
\begin{equation}\psi_T(a)\approx exp[\frac{\pi}{2GH^2}-\frac{3\pi a^2}{4G}]\end{equation}
Comparing Eqs.(19) and (29), we find that the only difference is
just an unimportant pre-exponential factor.
\par The
Hartle-Hawking(H-H) no boundary wave function is given by the path
integral
\begin{equation}\psi_{HH}=\int[dg][d\varphi]e^{-S_E(g,\varphi)}\end{equation}
In order to determine $\psi_{HH}$, we assume that the dominant
contribution to the path integral is given by the stationary
points of the action(the wormhole instantons) and evaluates
$\psi_{HH}$ simply as $\psi_{HH}\propto
e^{-S_E\mid_{saddle\_point}}$. When $(\dot{\varphi})^2\sim
\frac{1}{\lambda}$, from action(3) we can obtain
\begin{equation}S=\int\frac{3\pi}{4G}[(1-\dot{a}^2)adt-\int2\pi^2a^3H^2dt]\end{equation}
where $H^2=\frac{8\pi}{3}(v-\frac{1}{\lambda})$. The corresponding
Euclidean action $S_E=-i(S)_{continue}$ is
\begin{equation}S_E=\int-\frac{3\pi}{4G}[1+(\frac{da}{d\tau})^2]ad\tau+\int2\pi^2a^3H^2d\tau\end{equation}
where $\tau=it$. From action(31), we can obtain that the $a(t)$
satisfies the following classical equation of motion
\begin{equation}-(\frac{da}{dt})^2-1+H^2a^2=0\end{equation}
The solution of Eq.(33) is the de sitter space with
$a(t)=H^{-1}cosh(Ht)$. The corresponding Euclidean version
(replacing $t\rightarrow гн-i\tau$) of Eq.(33) is
\begin{equation}(\frac{da}{d\tau})^2-1+H^2a^2=0\end{equation}
The solution of Eq.(34) is
\begin{equation}a(\tau)=H^{-1}sin(H\tau)\end{equation}
Using Eqs.(32)(35), we obtain
\begin{equation}\psi_{HH}(a)\propto exp[-\frac{\pi}{2GH^2}(1-H^2a^2)^{\frac{3}{2}}]\end{equation}
The only difference between the H-H's wave function(36) and
Vilenkin's wave function(27) is the sign of the exponential factor.
The H-H's wave function(36) gives the probability distribution
\begin{equation}P_{HH}\propto e^{\frac{\pi}{2GH^2}}\end{equation}
The H-H's distribution(37) is the same as Vilenkin's one(28),
except a sign of the exponential factor. The distribution(37) is
peaked at $V-(1/\lambda)=0$ and it predicts a universe with a
positive cosmological constant $\frac{1}{\lambda}$. This is a very
significative result, it is different from zero cosmological
constant predicted by Coleman and Hawking.
\par When $V=\frac{1}{\lambda}$ for
our lagrangian, it represents a universe filled with Chaplygin
gas[5]. In fact H-H's wave function predicts a unverse filled with
Chaplygin gas. Next, we will discuss dark energy model of B-I type
scalar field and the relationship between our Lagrangian and
Chaplygin gas. \vskip 0.3 in \textbf{III.Dark Energy Model of B-I
Type Scalar Field}
\\
\\
\textsf{1. The Model With Lagrangian
$\frac{1}{\lambda}[1-\sqrt{1-\lambda
g^{\mu\nu}\varphi_{,\mu}\varphi_{,\nu}}]-V$}
\\
\\
In the spatially flat R-W metric, Einstein equation
$G_{\mu\nu}=KT_{\mu\nu}$ can be written as
\begin{equation}(\frac{\dot{a}}{a})^2=\frac{K}{3}T_0^0=\frac{K}{3}\rho\end{equation}
\begin{equation}2\frac{\ddot{a}}{a}+(\frac{\dot{a}}{a})^2=KT_1^1=KT_2^2=KT_3^3=-Kp\end{equation}
Substituting Eq.(38) into Eq.(39), we get
\begin{equation}\frac{\ddot{a}}{a}=-\frac{K}{6}(T_0^0-3T_1^1)\end{equation}
where
\begin{equation}T^\mu_\nu=\frac{g^{\mu\rho}\varphi_{,\nu}\varphi_{,\rho}}{\sqrt{1-\eta g^{\mu\nu}\varphi_{,\nu}\varphi_{,\rho}}}-\delta^\mu_\nu L_s\end{equation}
The energy density $\rho_s=T^0_0$ and pressure
$p_s=-T_1^1=-T_2^2=-T_3^3$ are defined as following
\begin{equation}\rho_s=V-\frac{1}{\lambda}+\frac{1}{\lambda\sqrt{1-\lambda\dot{\varphi}^2}}=\frac{B+\sqrt{1+\lambda c^2a^{-6}}}{\lambda}\end{equation}
\begin{equation}p_s=\frac{1}{\lambda}[1-\sqrt{1-\lambda\dot{\varphi}^2}]-V\end{equation}
where the upper index "." denotes the derivative with respect to
$t$, and $V(\varphi)$ is taken $V_0$ as a constant. We define
$B=\lambda V_0-1$. When $a(t)=0$ the kinetic energy
$\frac{\dot{\varphi}^2}{2}=\frac{1}{2\lambda}$ is critical value
from Eq.(5). According to Eqs.(38) and (42) we get
\begin{equation}(\frac{\dot{a}}{a})^2=\frac{K}{3\lambda}[\sqrt{1+\lambda c^2a^{-6}}+B]\end{equation}
When $a(t)$ is very small($a(t)\rightarrow0$), Eq.(44) approximates
to
\begin{equation}\dot{a}=\sqrt{\frac{Kc}{3\lambda^{\frac{1}{2}}a}}\end{equation}
\begin{equation}a\propto t^{\frac{2}{3}}\end{equation}
When $a(t)$ increases little by little until $a\gg(\lambda
c^2)^{\frac{1}{6}}$, Einstein equation(44) becomes
\begin{equation}(\frac{\dot{a}}{a})^2=\frac{K}{3}[V_0+\frac{1}{2}c^2a^{-6}]\end{equation}
From Eq.(47) we obtain
\begin{equation}\ddot{a}=\frac{KV_0}{3}a-\frac{Kc^2}{3}a^{-5}\end{equation}
From the above equation we find that when
$a>(\frac{c^2}{V_0})^{\frac{1}{6}}$, the universe is undergoing a
accelerated phase. Integrating Eq.(47) we obtain
\begin{equation}a^3\propto\sqrt{\frac{c^2}{2V_0}}\frac{(e^{\sqrt{12KV_0}t}-1)}{e^{\sqrt{3KV_0}t}}\end{equation}
When $a\rightarrow\infty$, Eq.(44) becomes
\begin{equation}(\frac{\dot{a}}{a})^2=\frac{K}{3}V_0\end{equation}
From the above equation, we see that the universe is undergoing a
inflation phase. However, the dark energy model interpolates
between a dust dominated phase in the past and a de-sitter
 phase at the late time.
\par Especially when $V_0=\frac{1}{\lambda}$, our
Lagrangian describes the Chaplygin gas that was proposed as a
model for both dark energy and dark matter in the present
universe[5]. Correspondingly, the density and pressure are
\begin{equation}\rho_c=\frac{\sqrt{1+\lambda c^2a^{-6}}}{\lambda}\end{equation}
\begin{equation}p_c=-\frac{1}{\rho_c\lambda^2}\end{equation}
Substituting Eq.(51) into Eq.(38), we obtain the solution of
Einstein equation(38)
\begin{equation}t=\frac{\lambda^{\frac{1}{2}}}{6}[ln\frac{(\frac{1}{\lambda^2}+\frac{c^2}{\lambda}a^{-6})^{\frac{1}{4}}+(\frac{1}{\lambda})^{\frac{1}{2}}}{(\frac{1}{\lambda^2}+\frac{c^2}{\lambda}a^{-6})^{\frac{1}{4}}-(\frac{1}{\lambda})^{\frac{1}{2}}}-2\arctan(\sqrt{1+c^2\lambda a^{-6}})^{\frac{1}{4}}]\end{equation}
When $a(t)$ is very small($a(t)\rightarrow0$), the density is
approximated by
\begin{equation}\rho_c\sim\frac{c}{a^3\lambda^{\frac{1}{2}}}\end{equation}
that corresponds to a universe dominated by dust-like matter. For
large values of the cosmological scalar factor $a(t)$ it follows
that
\begin{equation}\rho_c\sim\frac{1}{\lambda}\end{equation}
\begin{equation}p_c\sim-\frac{1}{\lambda}\end{equation}
which in turn corresponds to a universe with a cosmological
constant $\frac{1}{\lambda}$(i.e., a de-sitter universe). At early
time, i.e., the cosmological scale factor $a(t)$ is small,
$\rho_c\sim\frac{c}{a^3\lambda^{\frac{1}{2}}}$, which corresponds
to a dust like dominated universe. At late time, i.e., the
cosmological scale factor is large, $\rho_c\sim\frac{1}{\lambda}$,
which corresponds to a cosmological constant like dominated
universe. Therefore the Chaplygin gas interpolates between a dust
dominated phase in the past and a de-Sitter phase at late time.
\par When potential is taken to zero, the density and pressure are
\begin{equation}\rho_s=\frac{\dot{\varphi}^2}{\sqrt{1-\lambda\dot{\varphi}^2}}-\frac{1}{\lambda}[1-\sqrt{1-\lambda\dot{\varphi}^2}]\end{equation}
\begin{equation}p_s=\frac{1}{\lambda}[1-\sqrt{1-\lambda\dot{\varphi}^2}]\end{equation}
Substituting Eq.(5) to the above two expressions, we have
\begin{equation}w=\frac{p_s}{\rho_s}=\frac{a^3}{\sqrt{a^6+\lambda c^2}}\end{equation}
and can see
\begin{equation}0\leq w<1\end{equation}
So, there is no accelerated expansion in the universe of B-I type
scalar field without potential. For B-I type scalar field with
potential, we have
\begin{equation}\rho+3p=\frac{2}{\lambda}+\frac{3\lambda\dot{\varphi}^2-2}{\lambda\sqrt{1-\lambda\dot{\varphi}^2}}-2V(\varphi)\end{equation}
When potential is greater than $\frac{1}{\lambda}$ and the kinetic
energy of $\varphi$ field evolves to region of
$\dot{\varphi}^2<\frac{2}{3\lambda}$, $\rho+3p<0$. The universe
undergoes a phase of accelerating expansion.
\par We also get
\begin{equation}\rho+p=\frac{\dot{\varphi}^2}{\sqrt{1-\lambda\dot{\varphi}^2}}>0\end{equation}
and
\begin{equation}w=\frac{p}{\rho}>-1\end{equation}
However, some analysis to the observation date hold that the range
of state parameter lies in $-1.32<$w$<-0.82$[6]. From Eq.(63) we
know that parameter of state equation is larger than -1. In order
to give a favor explanation to the observation results, we
investigate the phantom field model that possess negative kinetic
energy and can realize w$<-1$ in their evolution. Next we discuss
phantom field model.
\\
\\
\textsf{2. The Model With Lagrangian
$\frac{1}{\lambda}[1-\sqrt{1+\lambda
g^{\mu\nu}\varphi_{,\mu}\varphi_{,\nu}}]-V(\varphi)$}
\\
\\
We consider the case that the kinetic energy terms is negative.
The energy-momentum tensor is
\begin{equation}T_\nu^\mu=-\frac{g^{\mu\rho}\varphi_{,\nu}\varphi_{,\rho}}{\sqrt{1+\lambda g^{\mu\nu}\varphi_{,\mu}\varphi_{,\nu}}}-\delta^\mu_\nu L\end{equation}
From Eq.(64), we have
\begin{equation}\rho=T_0^0=\frac{1}{\lambda\sqrt{1+\lambda\dot{\varphi}^2}}-\frac{1}{\lambda}+V\end{equation}
\begin{equation}p=-T_1^1--T_2^2=-T_3^3=\frac{1}{\lambda}-\frac{\sqrt{1+\lambda\dot{\varphi}^2}}{\lambda}-V\end{equation}
Based on Eq.(65) and (66), we can obtain
\begin{equation}\rho+p=-\frac{\dot{\varphi}^2}{\sqrt{1+\lambda\dot{\varphi}^2}}\end{equation}
It is clear that the static equation w$<-1$ is completely decided by
Eq.(67). We can also get
\begin{equation}\rho+3p=\frac{2}{\lambda}-\frac{2}{\lambda}\sqrt{1+\lambda\dot{\varphi}^2}-\frac{\dot{\varphi}^2}{1+\lambda\dot{\varphi}^2}-2V\end{equation}
It is obvious that $\rho+3p<0$. Eq.(68) shows that the universe is
undergoing a phase of accelerated expansion. The model of phantom
B-I type scalar field without potential $V(\varphi)$ is hard to
understand. In this model we can always find
$\rho=\frac{1}{\lambda\sqrt{1+\lambda\dot{\varphi}^2}}-\frac{1}{\lambda}<0$
and $(\frac{\dot{a}}{a})^2<0$. It is unreasonable apparently. In
the model of phantom B-I type scalar field with potential
$V(\varphi)$, if
$V(\varphi)>\frac{1}{\lambda}-\frac{1}{\lambda\sqrt{1+\lambda\dot{\varphi}^2}}$,
$\rho$ is greater than zero. \\
\par First, we consider the case of a specific simple example
$V=u_0=const$ and $u_0-\frac{1}{\lambda}=\frac{A}{\lambda}(A>0)$.
So, Eq.(63) becomes
\begin{equation}\rho=\frac{1}{\lambda\sqrt{1+\lambda\dot{\varphi}^2}}+\frac{A}{\lambda}\end{equation}
Substituting Eq.(69) into Eq.(38), we have
\begin{equation}(\frac{\dot{a}}{a})^2=\frac{K}{3}[\frac{1}{\lambda\sqrt{1+\lambda\dot{\varphi}^2}}+\frac{A}{\lambda}]\end{equation}
We can obtain from the Euler-Lagrange equation (4)
\begin{equation}\dot{\varphi}=\frac{c}{\sqrt{a^6-\lambda c^2}}\end{equation}
where $c$ is integrate constant. Substituting Eq.(71) into Eq.(70),
we get
\begin{equation}\dot{a}=\sqrt{\frac{Ka^2}{3\lambda}[\sqrt{1-\lambda c^2a^{-6}}+A]}\end{equation}
From the above equation, we know the universe is nonsingular
because the minimum $a_{min}$ of scale factor is $(\lambda
c^2)^{\frac{1}{6}}$. When the universe scale factor approximates
to $a_{min}$, Eq.(72) becomes
\begin{equation}\dot{a}=\sqrt{\frac{KA}{3\lambda}}a\end{equation}
\begin{equation}a=e^{\sqrt{\frac{KA}{3\lambda}}t}\end{equation}
When $a\rightarrow\infty$, Eq.(72) becomes
\begin{equation}\dot{a}=\sqrt{\frac{K(A+1)}{3\lambda}}a\end{equation}
\begin{equation}a=e^{\sqrt{\frac{K(A+1)}{3\lambda}}t}\end{equation}
\par In phantom model with constant potential, the universe is
always undergoing a phase of inflation at very small and lager
scale factor.
\par Next we study the cosmological evolution by
numerical analysis in the phantom model with a specific
potential[11]
 \begin{equation}V(\varphi)=V_0(1+\frac{\varphi}{\varphi_0})e^{(-\frac{\varphi}{\varphi_0})}\end{equation}
 \par As we consider the phantom field becomes dominant
, we can neglect the nonrelativitic and relativistic components
(matter and radiation) in the universe, then from Eqs.(5,38), we
have
\begin{equation}\ddot{\varphi}+\dot{\varphi}(1+\lambda\dot{\varphi}^2)\sqrt{3K[\frac{1}{\lambda\sqrt{1+\lambda\dot{\varphi}^2}}-\frac{1}{\lambda}+V(\varphi)]}-V^{'}(\varphi)(1+\lambda\dot{\varphi}^2)^\frac{3}{2}=0\end{equation}
where the overdot denotes the differentiation with respect to $t$
and the prime denotes the differentiation with respect to
$\varphi$.
\par To study an numerical computation, it is
convenient to introduce two independent variables
\begin{equation}\left\{\begin{array}{l}
   X=\varphi\\
   Y=\dot{\varphi}\end{array}\right.
   \end{equation}
then Eq.(78) can be written
\begin{equation}\left\{\begin{array}{l}\frac{dX}{dt}=Y \\
   \frac{dY}{dt}=V^{'}(X)(1+\lambda Y^2)^\frac{3}{2}-Y (1+\lambda Y^2)\sqrt{3K[\frac{1}{\lambda\sqrt{1+\lambda
   Y^2}}-\frac{1}{\lambda}+V(X)]}\end{array}\right.
   \end{equation}
   We can obtain this system's critical point from
\begin{equation}\left\{\begin{array}{l}\frac{dX}{dt}=0 \\
                                                  \frac{dY}{dt}=0
                                                  \end{array}
                                                  \right.
                                                  \end{equation}
   then its critical point is $(X_c,0)$, where the critical value $X_c$ is determined by $V'(X_c)=0$.
   Linearizing  Eq.(80) around the critical point, we have
   \begin{equation} \left\{\begin{array}{l}\frac{dX}{dt}=Y \\
                                                  \frac{dY}{dt}=V''(X_c)(X-X_c)-\sqrt{(3K V(X_c)}Y
                                                  \end{array}
                                                  \right.
                                                  \end{equation}
 the types of the critical point are determined by the eigenequation  of
 system
 \begin{equation}\varepsilon^2+\alpha\varepsilon+\beta=0\end{equation}
where $\alpha=\sqrt{3K V(X_c)}$,$\beta=-V''(X_c)$, the two
eigenvalues are $\varepsilon_1=\frac{-\sqrt{3K V(X_c)}+\sqrt{3K
V(X_c)+4V''(X_c)}}{2}$, $\varepsilon_2=\frac{-\sqrt{3K
V(X_c)}-\sqrt{3K V(X_c)+4V''(X_c)}}{2}$. For a positive potentials,
if $V''(X_c)<0$, then the critical point ($X_c$,0) is a stable node,
which implies that the dynamical system admits attractor solutions.
We can also conclude that if a potential possesses the general
properties: $V(X_c)>0,V'(X_c)=0$ and $ V''(X_c)<0$, then our phantom
model with this potential will have a attractor solution and predict
a late time de-sitter like behavior($w_\varphi=-1$).
 Substituting Eq.(77) into Eq.(80), we obtain
 \begin{equation}\left\{\begin{array}{l}\frac{dX}{dt}=Y \\
  \frac{dY}{dt}=-\frac{V_0}{\varphi^2_0}X e^{(-\frac{X}{\varphi_0})}(1+\lambda
   Y^2)^\frac{3}{2}
     -Y (1+\lambda Y^2)\sqrt{3K[\frac{1}{\eta\sqrt{1+\lambda
        Y^2}}-\frac{1}{\lambda}+V_0(1+\frac{X}{\varphi_0})e^{(\frac{X}{-\varphi_0})}]}
           \end{array}
            \right.
             \end{equation}
  \par From Eqs(81,84) we obtain the critical $X_c=0$ and
 $V''(X_c)=-\frac{V_0}{\varphi^2_0}<0$ from Eq(77). Therefore
 this model has an attractor solution which corresponds to its
 attractor regime, the equation of state $w\leq-1$.
 To solve this equations system via the numerical approach, we re-scale the quantities as
 $x=\frac{X}{\varphi_0}$, $s=(\varphi^2_0\lambda)^{-\frac{1}{2}}t$, $y=\sqrt{\lambda
 }Y$. Then Eq.(84) becomes \begin{equation}\left\{\begin{array}{l}\frac{dx}{ds}=y \\
                                                  \frac{dy}{ds}=-\gamma x e^{(-x)}(1+
                                                  y^2)^\frac{3}{2}
                                                  -\phi_0 y(1+y^2)\sqrt{3K[\frac{1}{\sqrt{1+
                                                  y^2}}-1+\gamma(1+x)e^{(-x)}]}
                                                  \end{array}
                                                  \right.
                                                  \end{equation}
 where we set $K=1$ and $\gamma=V_0\lambda$
 is parameter. The numerical results with different initial condition are plotted in
 Figs.$1-3$ and the parameters $\varphi_0=\sqrt{0.1}, \gamma=3$.
\begin{center}\vspace{0.5cm}
\includegraphics[angle=270,width=10cm]{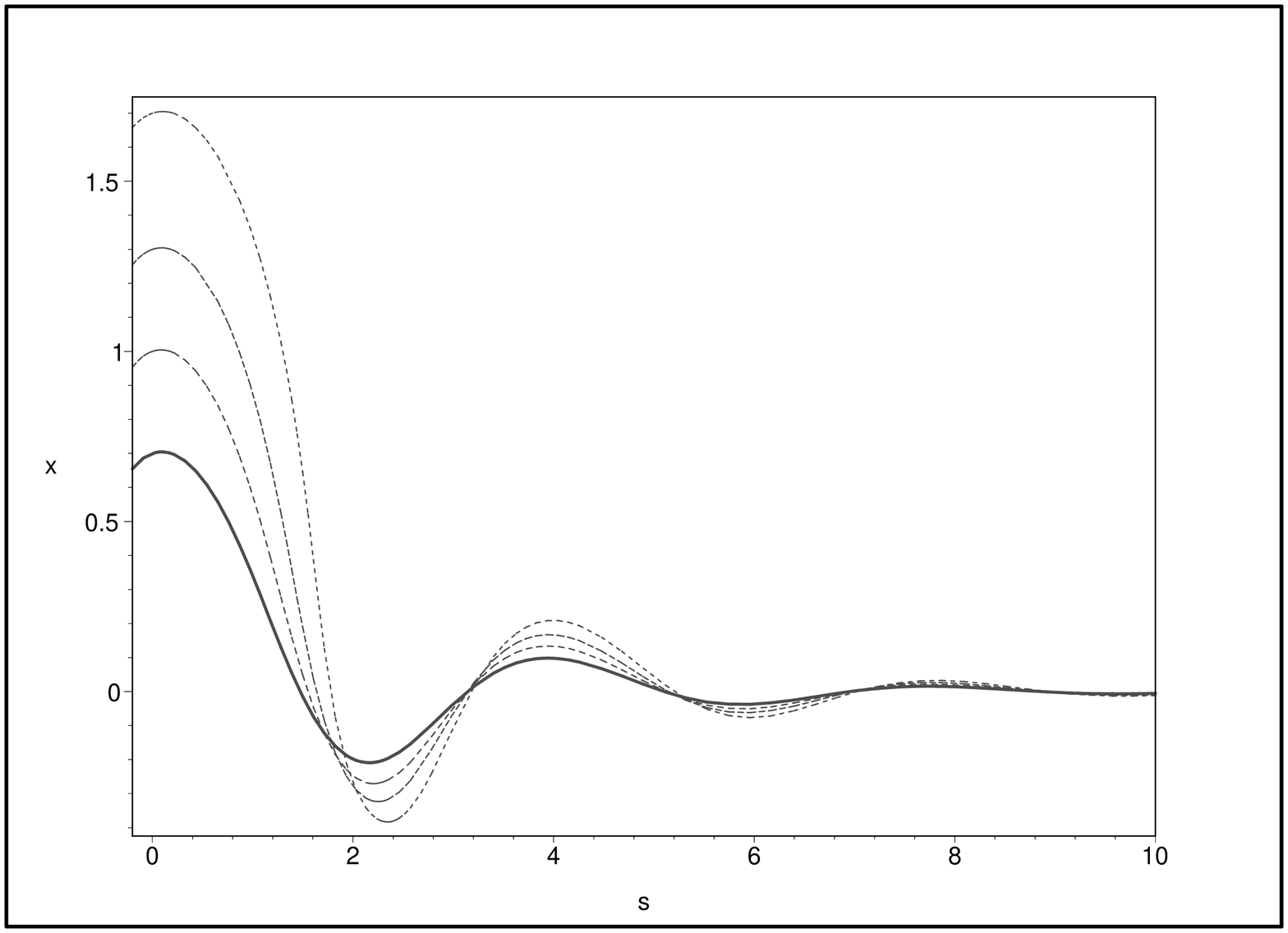} \end{center} \hfill
~\begin{minipage}{5.5in}~Fig1.This plot shows the evolution of the
scalar field in difference initial condition, solid line is for
$\varphi_{in}=0.7\varphi_0$, dotted line is for
$\varphi_{in}=\varphi_0$, dashed line and dot-dashed line for
$\varphi_{in}=1.3\varphi_0, 1.7\varphi_0$ respectively, they are
all plotted for a fixed value of $y_{in}=0.1$ .\end {minipage}
\begin{center}\vspace{0.5cm}
\includegraphics[angle=270,width=10cm]{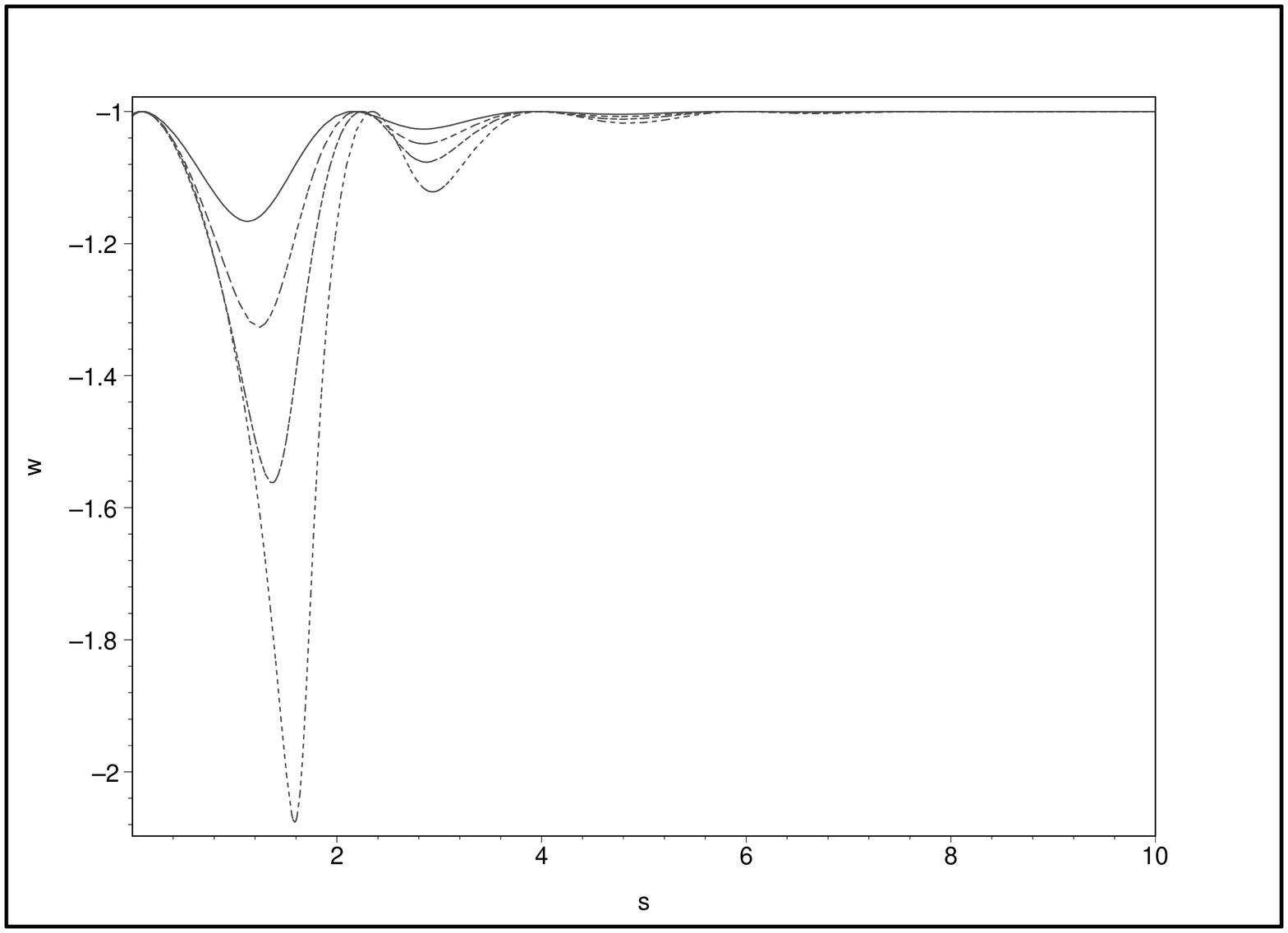} \end{center} \hfill
~\begin{minipage}{5.5in}~Fig2. The evolution of w with respect to
s, the initial condition is the same as fig1.\end {minipage}
\begin{center}\vspace{0.5cm}
\includegraphics[angle=270,width=10cm]{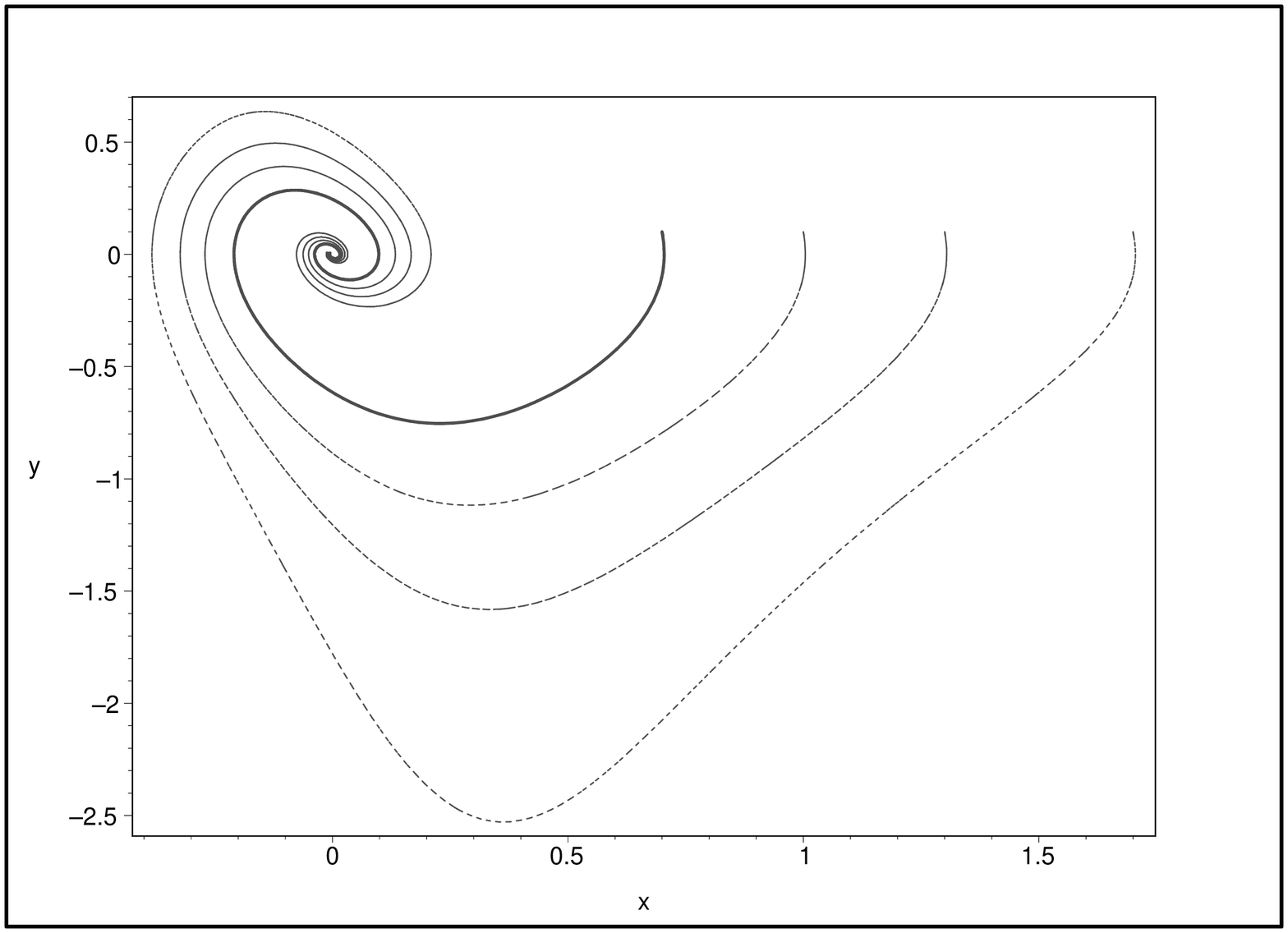} \end{center} \hfill
\begin{minipage}{5.5in}~Fig3. The attractor property of the
system in the phase plane, the initial condition is the same as
fig1.
\end{minipage}
\\
\par As we know,when $\lambda\rightarrow0$, our model turns to be phantom quintessence model. In order to see the nonlinear effect,  we  plot the
phantom quintessence model with our model in fig4 and fig5.
 \begin{center}\vspace{0.5cm}
\includegraphics[angle=270,width=10cm]{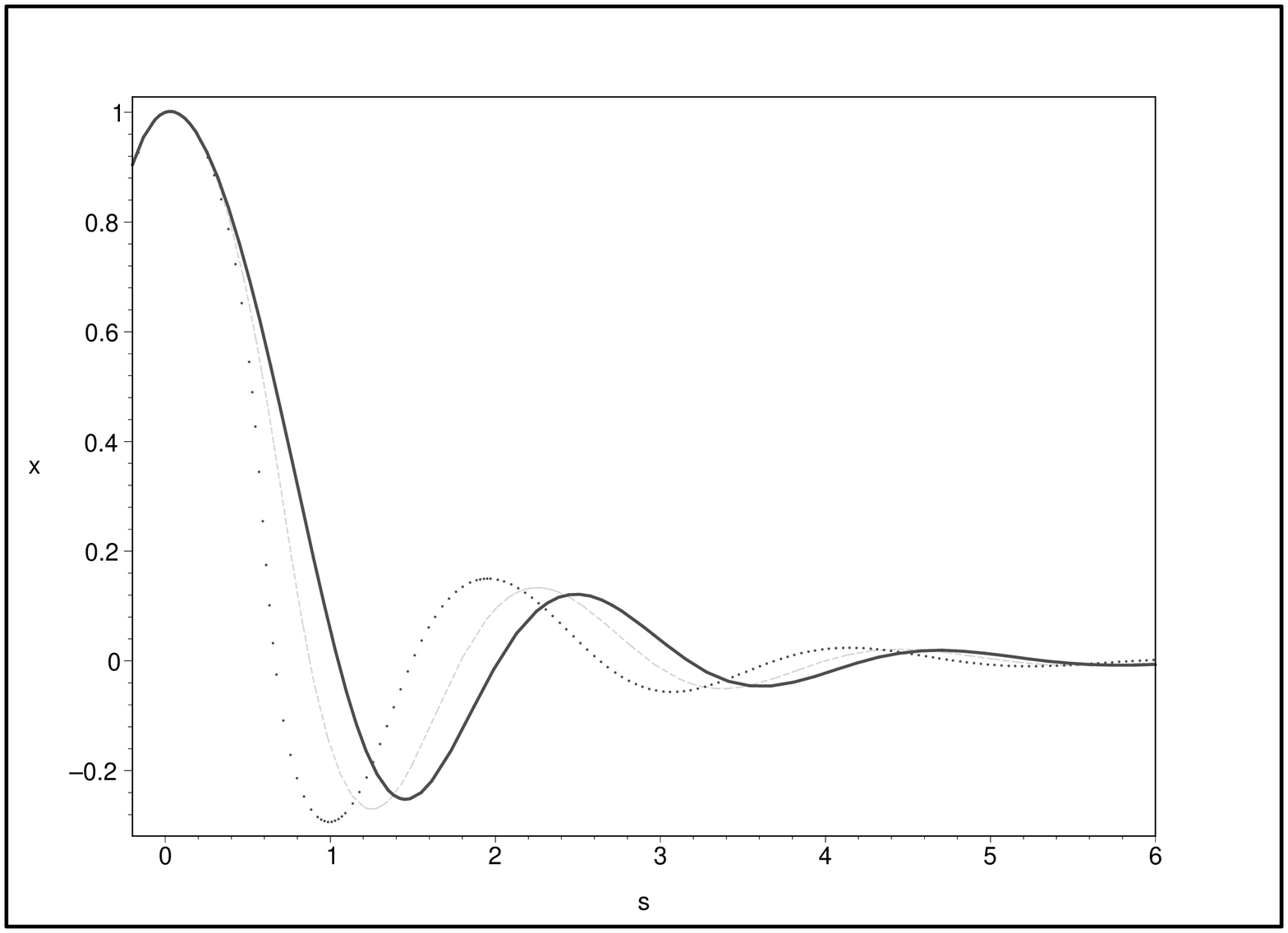} \end{center} \hfill
\par\begin{minipage}{5.5in} Fig4. the evolution of scalar field with respect to s, solid line is
phantom quintessence model, dotted line and dashed line are
nonlinear B-I phantom field,dotted line is for
$\lambda=2/3$,dashed line is for $\lambda=1/3$.
\end{minipage}
\begin{center}\vspace{0.5cm}
\includegraphics[angle=270,width=10cm]{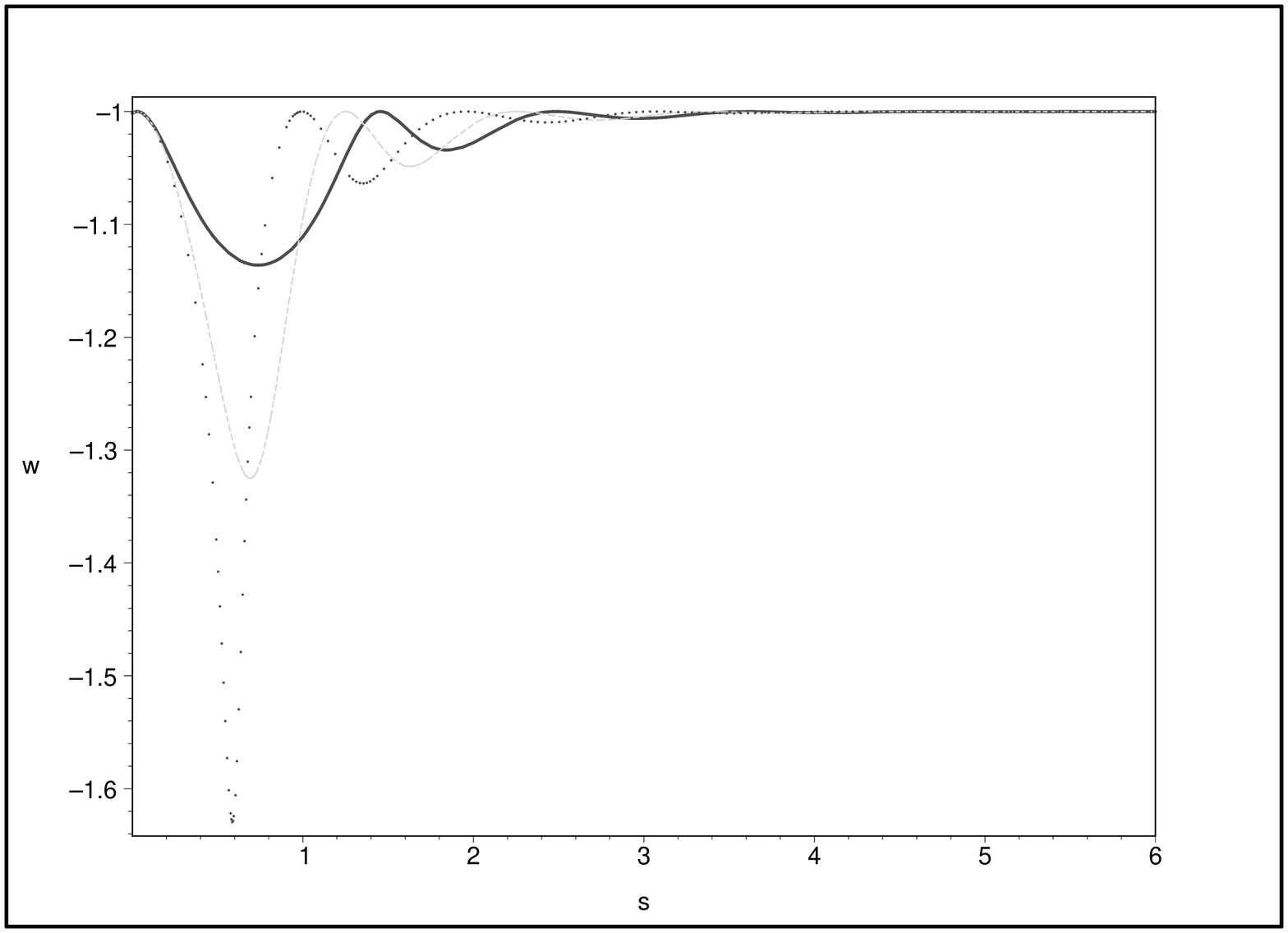} \end{center} \hfill
~\begin{minipage}{5.5in} Fig5.the evolution of w with respect to
s, solid line is phantom quintessence model, dotted line and
dashed line are nonlinear  B-I phantom field, dotted line is for
$\lambda=2/3$, dashed line is for $\lambda=1/3$.\end{minipage}
\par From fig1,fig2 and fig3, we can easily find the system admits
a attractor solution. The equation of state parameter w starts from
regime of near -1, then quickly evolves to the regime of smaller
than -1, last, turns back to execute the damped oscillation, and
reaches to -1 eventually for ever. Due to the unusual physical
properties in phantom model, the phantom field, releases at the
distance from the origin with a small kinetic energy, and moves
towards the top of the potential and crosses over to the other side,
then turns back to execute the damped oscillation around the
critical point. After a certain time the motion ceases and the
phantom field settles on the top of the potential permanently to
mimic the de Sitter-like behavior(w$_{\phi}=-1$). Fig4 and fig5
indeed shows that the nonlinear scalar phantom field will reduce to
phantom quintessence model when $\lambda$ decreases to zero. The
nonlinear effect does not affect the global attractor behavior but
change the evolution of w and scalar field $\phi$ in details.

\begin{center}\textbf{IV.CONCLUSION}\end{center}
\par It shows that the tachyon can be described by a
B-I type Lagrangian resulting from string theory. It is clear that
our Lagrangian $\frac{1}{\lambda}[1-\sqrt{1-\lambda
g^{\mu\nu}\varphi_{,\mu}\varphi_{,\nu}}]-V(\varphi)$ is equivalent
formally to the tachyon type Lagrangian
$-\frac{1}{\lambda}\sqrt{1-g^{\mu\nu}\Phi_{,\mu}\Phi_{,\nu}}$, with
a potential $\frac{1}{\lambda}-V(\varphi)$ where we re-scale the
scalar field as $\Phi=\lambda^{\frac{1}{2}}\varphi$. The WD equation
are solved analytically for both very large and small $a(t)$. For
the very small cosmological scale factor, the Vilenkin's tunnelling
wave function Eq.(27) predicts a nucleating universe with the
biggest possible vacuum energy $V$(cosmological constant) and the
smallest possible critical kinetic energy $\frac{1}{2\lambda}$. The
only difference between the H-H's wave function Eq.(36) and
Vilenkin's wave function Eq.(27) is the sign of the exponential
factor. Then the H-H's probability distribution in Eq.(37) reaches
to peak at $V-\frac{1}{\lambda}=0$. It predicts a universe with a
postive cosmological constant $\frac{1}{\lambda}$. It is different
from zero cosmological constant that has been predicted by Coleman
and Hawking.
\par The parameter w of state equation lies in the range of
$0\leq$w$<1$ for B-I type scalar field without potential. The
universe of B-I type scalar field with potential can undergo a
phase of accelerated expansion. The corresponding parameter of
state equation lies in the range of $-1<$w$<-\frac{1}{3}$. This
model admits a late time attractor solution that leads to an
equation of state w$=-1$.
\par In order to give a favor explanation to the observational
results that the range of state parameter lies in
$-1.32<$w$<-0.82$, we investigate the phantom field model that
possess negative kinetic energy.
 The weak and strong energy
condition are violated for phantom B-I type scalar field. The
parameter w of state equation lies in the range of w$<-1$. When
the potential $V(\varphi)=u_0$, the universe is nonsingular and
stays in de-sitter phase always. When we choose a potential as
Eq.(77), the evolution behavior of the state parameter w is shown
in Fig.2. By numerical analysis, we learn that there is an
attractor solution in Fig.3.
\\
 \textbf{ACKNOWLEDGEMENTS}
\\
\par This work is partly supported by NNSFC under Grant No.10573012 and No.10575068 and by Shanghai
 Municipal Science and Technology Commission No.04dz05905.
\\
\\
{\noindent\Large \bf References} \small{
\begin{description}
\item {[1]}{W. Heisenberg, Z.Phys, \textbf{133}, 79(1952); \textbf{126}, 519(1949); \textbf{113}, 61(1939).}
\item {[2]}{M. Born and Z. Infeld, Proc. Roy. Soc. A \textbf{144}, 425(1934).}
\item {[3]}{H.P. de Oliveira, J. Math. Phys, \textbf{36}, 2988(1995);
\\I. Janis, E.T. Newman and J. Winicour, Phys. Rev. Lett, \textbf{20}, 878(1968);
\\T. Taniuti, Prog. Theor. Phys. (Kyoto) Suppl. \textbf{9}, 69(1958).}
\item {[4]}{A. Vilenkin, Phys. Rev. D \textbf{37}, 888(1988);
\\A. Vilenkin, Phys.Rev.D \textbf{50}, 2581(1994);
\\A. Vilenkin, gr-qc/9804051;gr-qc/9812027;
\\J.B. Hartle and S.W. Hawking, Phys. Rev. D \textbf{28}, 2960(1983);
\\L.P. Grishchuk and L.V. Rozhangsky, Phys. Lett. B \textbf{208}, 369(1988);
\\L.P. Grishchuk and L.V. Rozhangsky, Phys. Lett. B \textbf{234}, 9(1990);
\\A. Lukas, Phys. Lett. B \textbf{347}, 13(1995);
\\T. Vachaspati and A. Vilenkin, Phys.Rev. D \textbf{37}, 898(1988);
\\S.W. Hawking and N.G. Turok, gr-qc/9802062.}
\item {[5]}{A. Kamenshchik, U. Moschella, V. Pasquier, Phys. Lett. B \textbf{511}, 265(2001);
\\N.Bilic, G.B. Tupper, R.D. Viollier, Phys. Lett. B \textbf{535}, 17(2002);
\\R. Collister, J.C. Fabris, S.V.B. Goncalves, P.E. de Souza,
gr-qc/0210079.}
\item {[6]} {R.R. Caldwell, Phys. Rev. Lett. B \textbf{545}, 23(2002);\\
            V. Faraoni, Int. J. Mod. Phys. D \textbf{11}, 471(2002);\\
            S. Nojiri and S. D. dintsov, hep-th/0304131;hep-th/0306212;\\
            E. schulz and M. White, Phys. Rev. D \textbf{64}, 043514(2001);\\
            T. Stachowiak and Szydllowski, hep-th/0307128;\\
            G.W. Gibbons, hep-th/0302199;\\
            A. Feinstein and S. Jhingan, hep-th/0304069;\\
            A. Melchiorri, astro-ph/0406652.}
\item {[7]}{J.M. Aguirregabiria, L.P. Chiemento and R. Lazkoz, Phys. Rev. D \textbf{70}, 023509(2004).}
\item {[8]}{L. P. Chimento, Phys. Rev. D \textbf{76}, 123517(2004).}
\item {[9]}{M. Sami, Mod. Phys. Lett. A \textbf{19}, 1509(2004)\\
            P. Singh, M. Sami, N. Dadhich,Phys. Rev. D \textbf{68},023522(2003)\\
            D.J. Liu and X.Z. Li, Phys. Rev. D \textbf{68}, 067301(2003);\\
            M.P. Dabrowski et.al., Phys. Rev. D \textbf{68}, 103519(2003);\\
            S.M. Carroll, M. Hoffman, M. Teodden, astro-th/0301273;\\
            Y.S. Piao, R.G. Cai, X.M. Zhang and Y.Z. Zhang, hep-ph/0207143;\\
            J.G. Hao and X.Z. Li, hep-th/0305207;\\
            S. Mukohyama, Phys. Rev. D \textbf{66}, 024009(2002);\\
            T. Padmanabhan, Phys. Rev. D \textbf{66}, 021301(2002);\\
            M. Sami and T. Padamanabhan, Phys. Rev. D \textbf{67}, 083509(2003);\\
            G. Shiu and I. Wasserman, Phys. Lett. B \textbf{541}, 6(2002);\\
            L. kofman and A. Linde, hep-th/020512;\\
            H.B. Benaoum, hep-th/0205140;\\
            L. Ishida and S. Uehara, hep-th/0206102;\\
            T. Chiba, astro-ph/0206298;\\
            T. Mehen and B. Wecht, hep-th/0206212;\\
            A. Sen, hep-th/0207105;\\
            N. Moeller and B. Zwiebach, JHEP\textbf{0210}, 034(2002);\\
            J.M. Cline, H. Firouzjahi and P. Martineau, hep-th/0207156;\\
            S. Mukohyama, hep-th/0208094;\\
            P. Mukhopadhyay and A. Sen, hep-th/020814;\\
            T. Okunda and S. Sugimoto, hep-th/0208196;\\
            G. Gibbons, K. Hashimoto and P.Yi, hep-th/0209034;\\
            M.R. Garousi, hep-th/0209068;\\
            B. Chen, M. Li and F. Lin, hep-th/0209222;\\
            J. Luson, hep-th/0209255;\\
            C. kin, H.B. Kim and O.K. Kwon, hep-th/0301142;\\
            J.M. Cline, H. Firouzjahi and P. Muetineau, hep-th/0207156;\\
            G. Felder, L. Kofman and A. Starobinsky, JHEP\textbf{0209}, 026(2002);\\
            S. Mukohyama, hep-th/0208094;\\
            G.A. Diamandis, B.C. Georgalas, N.E. Mavromatos and E. Pantonopoulos, hep-th/0203241;\\
            G.A. Diamandis, B.C. Georgalas, N.E. Mavromatos, E. Pantonopoulos and I. Pappa, hep-th/0107124;\\
            M.C. Bento, O. Bertolami and A. Sen, hep-th/020812;\\
            H. Lee et.al., hep-th/0210221;\\
            M. Sami, P. Chingangbam and T. Qureshi, hep-th/0301140;\\
            F. Leblond and A.W. Peet, hep-th/0305059;\\
            J.G. Hao and X.Z. Li, Phys. Rev. D \textbf{66}, 087301(2002);\\
            X.Z. Li and X.H. Zhai, Phys. Rev. D \textbf{67}, 067501(2003);\\
            J.G. Hao and X.Z. Li, Phys. Rev. D \textbf{68}, 043501(2003);\\
            Burin Gumjudpai, Tapan Naskar, M. Sami and Shinji Tsujikawa,
hep-th/0502191;
\\Yi-Huan Wei, gr-qc/0502077, gr-qc/0410050,astro-ph/0405368;Shin'ichi Nojiri, Sergei D. Odintsov, Shinji Tsujikawa, Phys. Rev. D \textbf{71},
063004(2005);
\\Luis P. Chimento, Ruth Lazkoz, astro-ph/0411068;
\\ Pedro F. Gonzalez-Diaz,  Phys. Rev. D \textbf{70},
063530(2004);
\\Shin'ichi Nojiri, Sergei D.Odintsov, Phys. Rev. D \textbf{70}, 103522(2004);
\\Pedro F. Gonzalez-Diaz, Carmen L. Siguenza, Nucl. Phys. B
\textbf{697}, 363(2004);
\\ E. Babichev, V. Dokuchaev, Yu.
Eroshenko, Class.Quant.Grav. \textbf{22}, 143(2005);
\\Federico Piazza, Shinji Tsujikawa, JCAP \textbf{0407},
004(2004);
\\Yi-Huan Wei, Yu Tian, Class.Quant.Grav. \textbf{21}, 5347(2004);
\\Emilio Elizalde, Shin'ichi Nojiri, Sergei D. Odintsov, Phys. Rev. D \textbf{70},
043539(2004);
\\Mariam Bouhmadi-Lopez, Jose A. Jimenez Madrid, astro-ph/0404540;Jian-gang Hao, Xin-zhou Li, Phys. Lett. B \textbf{606},
7(2005);
\\Shin'ichi Nojiri, Sergei D. Odintsov,  Phys. Lett. B \textbf{599},
137(2004);
\\Pedro F. Gonzalez-Diaz, Phys.Rev. D\textbf{69}, 063522(2004);
\\Iver Brevik, Shin'ichi Nojiri, Sergei D. Odintsov, Luciano Vanzo, Phys. Rev. D \textbf{70},
043520(2004);
\\ Jian-gang Hao, Xin-zhou Li, Phys. Rev. D \textbf{70},
043529(2004);
\\Xin-zhou Li, Jian-gang Hao, Phys. Rev. D \textbf{69},  107303(2004).}
\item {[10]}{S.Weinberg, Rev.Mod.Phys \textbf{61}, 1(1989);
\\S.Coleman, 1988b, "Why is there nothing rather than something: A theory of the cosmological
constant", Harvard University Preprint No. HUTP-88/A022.}
\item {[11]} {E.Witten,Phys.Rev.D\textbf{46}5467(1992),Phys.Rev.D\textbf{47}3405(1993);\\
              K.Li and E.Witten, Phys.Rev.D\textbf{48}853(1993);\\
              D.Kutasov, M.Marino and G.Moore, JHEP\textbf{0010}045(2000);\\
              T.Chiba, astro-ph/0206298.}
\end{description}}
\end{document}